\documentclass[smallextended]{svjour3}     
\smartqed  
\usepackage{color}
\usepackage{ulem}
\usepackage{graphicx}
\usepackage[]{natbib}
\usepackage{amsmath,amsfonts}

\begin{document}


\title{Coupling between corotation and Lindblad resonances 
in the elliptic planar three-body problem}


\thanks{Grants or other notes
about the article that should go on the front page should be
placed here. General acknowledgments should be placed at the end of the article.}

\titlerunning{CoraLin}        


\author{Maryame El Moutamid \and Bruno Sicardy \and \\ St\'efan Renner
}

\institute{\textbf{Maryame El Moutamid}\at
              LESIA / IMCCE - Observatoire de Paris \\
              5, Place Jules Janssen \\
              92195 Meudon Cedex, France \\
              Tel.: +33-1-45077492 \\
              Fax: +33-1-45077110 \\
              \email{maryame.elmoutamid@obspm.fr}           
           \and
             \textbf{Bruno Sicardy} \at
              LESIA - Observatoire de Paris \\
              University Pierre et Marie Curie, Paris
           \and
             \textbf{St\'efan Renner} \at
             Laboratoire d'Astronomie de Lille Universit\'e Lille 1\\
             IMCCE - Observatoire de Paris           
}


\date{The date of receipt and acceptance should be inserted later. Version: \today}

\maketitle


\begin{abstract}
We investigate the dynamics of two satellites with masses $\mu_s$ and $\mu'_s$
orbiting a massive central planet in a common plane, 
near a first order mean motion resonance $m$+1:$m$ ($m$ integer).
We consider only the resonant terms of first order in eccentricity 
in the disturbing potential of the satellites, 
plus the secular terms causing the orbital apsidal precessions.
We obtain a two-degree of freedom system, associated with
the two critical resonant angles 
$\phi= (m+1)\lambda' -m\lambda - \varpi$ and 
$\phi'= (m+1)\lambda' -m\lambda - \varpi'$,
where $\lambda$ and $\varpi$ are the mean longitude and longitude of periapsis
of $\mu_s$, respectively, and where the primed quantities apply to $\mu'_s$.
We consider the special case where $\mu_s \rightarrow 0$ (restricted problem).
The symmetry between the two angles $\phi$ and $\phi'$ is then broken,
leading to two different kinds of resonances, classically referred to as 
Corotation Eccentric resonance (CER) and Lindblad Eccentric Resonance (LER), respectively.
We write the four reduced equations of motion near the CER and LER, 
that form what we call the CoraLin model. 
This model depends upon only two dimensionless parameters that control the dynamics of the system:
the distance $D$ between the CER and LER, and a forcing parameter $\epsilon_L$
that includes both the mass and the orbital eccentricity of the disturbing satellite.
Three regimes are found:  
for $D=0$ the system is integrable, 
for $D$ of order unity, it exhibits prominent chaotic regions,
while for $D$ large compared to 2, the behavior of the system is regular and can be qualitatively described 
using simple adiabatic invariant arguments.
We apply this model to three recently discovered small Saturnian satellites dynamically linked  
to Mimas through first order mean motion resonances : Aegaeon, Methone and Anthe.
Poincar\'e surfaces of section reveal the dynamical structure of each orbit, 
and  their proximity to  chaotic regions.
This work may be useful to explore various scenarii of resonant capture for those satellites.
\end{abstract}


\newpage

\section{Introduction}
\label{sec_intro}

We consider the problem of two small bodies of masses $\mu_s$ and $\mu'_s$ orbiting 
in a common plane around a central massive body of mass $M_{p}$ ($\mu_s, \mu'_s \ll M_{p}$).
In this paper, the central massive body will be called the planet, while the two orbiting objects
will be called the satellites\footnote{For sake of brevity, the masses 
$M_p$, $\mu_s$ and $\mu'_s$ will denote at the same time the bodies and their masses.}.
We consider a configuration that is close to a first-order $m+1$:$m$ mean motion resonance:
$$
(m+1)n' \approx mn,
$$ 
where $m$ is an integer (positive or negative depending on whether $\mu_s$ orbit inside or outside $\mu'_s$),
and $n$ and $n'$ are the mean motions of  $\mu_s$ and $\mu'_s$, respectively.
Near the resonance, the dynamics of the satellites is described by a two-degree of freedom system with
two critical resonant angles $\phi$ and $\phi'$:
\begin{equation}
\begin{array}{ll}
\phi =  & (m+1)\lambda' - m\lambda - \varpi \\
\phi' =  & (m+1)\lambda' - m\lambda - \varpi',
\end{array}
\label{eq_crit}
\end{equation}
where $\lambda$ and $\varpi$ and their primed counterparts are the classical notations for the 
mean longitude and longitude of periapsis of the satellites, respectively.

Having two degrees of freedom, the Hamiltonian describing the motion of $\mu_s$ and $\mu'_s$
is in general not integrable, and leads to chaotic behaviors in certain regions of phase space,
as shown herein.  

The aims of this paper are:
(1) To describe the problem by a generic way, also to rescale the restricted problem ($\mu_s=0$) 
so that it depends upon two dimensionless parameters only: 
the distance $D$ between the two resonances and 
a parameter $\epsilon_L$ that depends upon the mass and orbital eccentricity of $\mu'_s$, 
thus allowing a generic approach of the problem.
While the problem is integrable when the $D=0$,
we show numerically that large chaotic regions appear for small distances. 
For large $D$'s, the system tends again toward an integrable system that
can be solved using adiabatic invariance arguments.
(2) A second goal of this paper is to clearly distinguish the effects of the two kinds of resonances associated with $\phi$ and $\phi'$.
When $\mu_s \neq 0$ and $\mu'_s \neq 0$, the two resonances have indeed symmetric behaviors,
but this symmetry is broken when for instance $\mu_s=0$, that is, the planar restricted three-body problem.
The angle $\phi$ then describes the so-called Lindblad Eccentric Resonance (LER), 
while $\phi'$  describes the Corotation Eccentric Resonance (CER).
This terminology was originated from galactic dynamics, 
see e.g. \textcolor{blue}{\cite{lind61,lind62,lin64,gol79}}, 
and it's used in the case of planetary rings, however, it's not frequently used in Celestial Mechanics. 
As recalled later, LER's mainly excite orbital eccentricities
(leaving semi-major axes largely unaffected),  
while CER's mainly change semi-major axes 
(leaving eccentricities largely unaffected).
The final goal of this paper is (3) to discuss the integrability of the two-degree of freedom
system in the presence of the two critical angles $\phi$ and $\phi'$.
There is a considerable amount of literature for the case $\mu_s=0$ and $e'=0$ 
(the planar, restricted and circular three-body problem), in which only the critical angle $\phi$ 
appears, reducing the problem to a one-degree of freedom integrable system described by
the classical Andoyer Hamiltonian given in Eq.~(\ref{eq_H_And}), 
see \textcolor{blue}{\cite{hen83,fer85,fer07}}.
The problem where both $\phi$ and  $\phi'$ are present has been treated by \textcolor{blue}{\cite{ses84}}
in the Keplerian case (i.e. with a central potential $\propto -{\cal G}M_{p}/r$, where
${\cal G}$ is the gravitation constant and $r$ is the distance to $M_{p}$) and for
$\mu_s \neq 0$ and $\mu'_s \neq 0$.
These authors show that the two-degree of freedom system is then integrable. 
More precisely, they show that the problem can be reduced to a one-degree of freedom
system described again by an Andoyer Hamiltonian.
More discussion about this result and its developments is provided in Section~\ref{sec_const}.

Our work can be applied in a general way to various problems involving ring and satellite dynamics. 
This extends for instance the approach of \textcolor{blue}{\cite{gol86}} and \textcolor{blue}{\cite{por91}}, 
who proposed a model to explain the stability of Neptune's incomplete rings (arcs) under 
the combined effects of Lindblad and corotation resonances. 
More recently, \textcolor{blue}{\cite{coo08}} and \textcolor{blue}{\cite{hed09,hed10}} have studied 
the motion of the small Saturnian satellites Anthe, Aegaeon and Methone that
are trapped in corotation resonances with Mimas, while being perturbed by nearby Lindblad resonances.


\section{General case}

\subsection{Derivation of the Hamiltonian}
\label{sec_hal}

We use here standard notations: $a$, $e$, $n$, $\lambda$ and $\varpi$ denote the 
semi-major axis, orbital eccentricity, mean motion, mean longitude and longitude of periapsis
of $\mu_s$, respectively, with similar primed quantities for $\mu'_s$.
For an oblate planet, 
this elements denote
the geometric elements (and not the osculating elements), 
see \textcolor{blue}{\cite{bor94,ren06}} for details. 
In that case, the quantities $\dot{\varpi}_s$ and $\dot{\varpi}'_s$ will
denote the secular time variations of  $\varpi_s$ and $\varpi'_s$
(precession rates) arising from the planet oblateness,
i.e. the variations that are not due to the resonances themselves.

We assume that the semi-major axes of $\mu_s$ and $\mu'_s$ remain close to reference
values $a_0$ and $a'_0$, respectively, where the mean motions are $n_0$ and $n'_0$, 
with $n_0=\sqrt{GM_{p}/a^{3}_{0}}$ and $n'_0=\sqrt{GM_{p}/a'^{3}_{0}}$.
Those reference values are chosen in a uniquely defined way so that
$(m+1)n'_0 - mn_0 - \dot{\varpi}'_s=0$, and so that the total orbital energy
of $\mu_{s}$ and $\mu_{s}'$ is $-{\cal G}M_p\mu_s/2a_0 - {\cal G}M_p\mu'_s/2a'_0$.
Then 
$(m+1)n'_0 -mn_0 - \dot{\varpi}_s= \dot{\varpi}'_s - \dot{\varpi}_s$, 
which defines the distance (in term of frequency) between the two resonances.
This is more clearly evident by noting that or $\dot{\varpi}'_s - \dot{\varpi}_s = \dot{\phi} - \dot{\phi'}$.
This choice for the reference values $a_0$ and $a'_0$ is arbitrary, and is
motivated by the fact that we will study later the behavior of a test particle near the corotation
resonance, where $(m+1)n' - mn - \dot{\varpi}'_s=0$.
We define:
$$
\xi= \frac{a-a_0}{a_0},~~~ 
\xi'= \frac{a'-a'_0}{a'_0}, 
$$ 
as the relative differences of the semi-major axes with respect to the reference radii.
We finally assume that $\mu_s$ and $\mu'_s$ stay far apart, in the sense that  their
orbital eccentricities and excursions in semi-major axes are small compared to their 
relative orbital separation: 
$\xi, \xi', e,e' \ll |a-a'|/a' \sim 1/m$.

The expansion of the disturbing function acting on $\mu_s$ and $\mu'_s$ to first order in eccentricities
yields two terms slowly varying with time: 
${\cal G}\mu_s \mu_{s}' e A \cdot \cos(\phi)$ and ${\cal G}\mu_s \mu_{s}' e' A' \cdot \cos(\phi')$, where 
$\phi$ and $\phi'$ are given in Eq.~(\ref{eq_crit}).
Those terms include both direct and indirect parts from the perturbing function.
The quantities $A$ and $A'$ are combinations of Laplace coefficients $b_{1/2}^{(m)}$, 
see \textcolor{blue}{\cite{ell00}} for details.
For numerical purposes, it is useful to note that $A$ and $A'$ have opposite signs, and that
\begin{equation}
 \begin{array}{l}
\displaystyle
  A = A^{m}(\alpha) =  \frac{1}{2a'} \left[ 2(m+1) + \alpha {\cal D}\right]  b_{1/2}^{(m+1)} (\alpha) \approx +\frac{0.802m}{a'} \\ \\
\displaystyle
  A' = A'^{m}(\alpha) =  -\frac{1}{2a'} \left[ (2m+1) + \alpha {\cal D}\right]  b_{1/2}^{(m)} (\alpha) \approx -\frac{0.802m}{a'}, \\
 \end{array}
\label{eq_AA'}
\end{equation} 
where $\alpha= a/a'$, ${\cal D}= d/d\alpha$, and where the approximations hold in the case of large $m$'s.
 Note that $A$ and $A'$ have the dimension of the inverse of a distance.

The derivation of the Hamiltonian of the system is classical and is described in many works, 
see e.g. \textcolor{blue}{\cite{las95}} for a general approach. 
Averaging the rapidly varying terms to zero, and keeping only the terms containing $\phi$ and $\phi'$,
we obtain the averaged Hamiltonian:
\begin{equation}
{\cal H}_1= 
-\frac{\mu^3({\cal G}M)^2}{2\Lambda^2}
-\frac{\mu'^3({\cal G}M')^2}{2\Lambda'^2} +
{\cal G}\mu \mu'  A \sqrt{\frac{2\Gamma}{\Lambda}} \cdot \cos(\phi) + 
{\cal G}\mu \mu'  A' \sqrt{\frac{2\Gamma'}{\Lambda'}} \cdot \cos(\phi')
- \dot{\varpi}_s\Gamma - \dot{\varpi}'_s\Gamma'.
\label{eq_H1}
\end{equation}
If $\mu'_s$ orbits inside $\mu_s$, then 
$M= M_p(M_p+\mu_s+\mu'_s)/(M_p+\mu'_s)$, 
$M'= M_p^3/(M_p+\mu'_s)^2$, 
$\mu= \mu_s (\mu'_s+M_p)/(\mu_s+\mu'_s+M_p)$ and 
$\mu'= \mu'_s (\mu'_s+M_p)/M_p$, see  \textcolor{blue}{\cite{ses84}}.
Equivalent expressions may be derived if $\mu_s$ orbits inside $\mu'_s$.
In all cases, the orbital elements refer to the center of mass of $M_p$ and the innermost satellite.
Note that since $\mu_s, \mu'_s \ll M_{p}$,
we have
$M \approx M' \approx M_p, \mu \approx \mu_s$ and $\mu' \approx \mu'_s$.

Moreover, since the functional dependence of the state vectors of $\mu$ and $\mu'$ upon 
the geometric elements is the same as for Keplerian case, to first order in eccentricities, 
the values of $A$ and $A'$ can be directly derived from the formulae~(\ref{eq_AA'}),
provided geometric elements are used instead of osculating elements.

The actions $\Lambda$, $\Gamma$, $\Lambda'$ and $\Gamma'$ are the Poincar\'e variables, 
which are respectively conjugates to the angle variables 
$\lambda$, $-\varpi$, $\lambda'$  and $-\varpi'$ as shown below:
\begin{equation}
\begin{array}{rl}
\lambda  \longleftrightarrow & \Lambda = \mu\sqrt{{\cal G}Ma} \\
-\varpi     \longleftrightarrow & \Gamma= \mu\sqrt{{\cal G}Ma}(1 -\sqrt{1-e^2}) \approx  \mu e^2\sqrt{{\cal G}Ma}/2 \\
\lambda'  \longleftrightarrow & \Lambda' = \mu'\sqrt{{\cal G}M'a'} \\
-\varpi'     \longleftrightarrow & \Gamma'= \mu'\sqrt{{\cal G}M'a'}(1 -\sqrt{1-e'^2}) \approx  \mu' e'^2\sqrt{{\cal G}M'a'}/2.
\label{eq_Poin}
\end{array}
\end{equation}

Because ${\cal H}_1$ depends only on $\phi$ and $\phi'$, 
it is convenient to use the new pairs of conjugate variables:
\begin{equation}
\begin{array}{rl}
\lambda  \longleftrightarrow & J = \Lambda + m (\Gamma + \Gamma')         \\
\lambda'  \longleftrightarrow & J'=  \Lambda' - (m+1) (\Gamma + \Gamma')  \\
\phi     \longleftrightarrow & \Theta= \Gamma                                                     \\
\phi'     \longleftrightarrow & \Theta' = \Gamma',                                                 \\
\end{array}
\label{eq_newvar}
\end{equation}
and introduce the new actions $\Lambda$, $\Gamma$, $\Lambda'$ and $\Gamma'$
into ${\cal H}_1$.

\subsection{Physical interpretation of the actions}
\label{sec_const}

In order to better understand globally the motions of $\mu$ and $\mu'$, 
it is instructive to consider the various actions entering in the system.
Because ${\cal H}_1$ does not depends on $\lambda$ and $\lambda'$, 
$J$ and $J'$ are constants of motion.
Consequently, the initial four-degree of freedom system (two satellites moving in a common plane)
reduces to a two-degree of freedom system. 
It is generally not integrable (see e.g. Fig.~\ref{sum}), unless $\dot{\varpi}_s - \dot{\varpi}'_s=0$, 
as discussed later.

Turning back to $J$ and $J'$, we have:
\begin{equation}
\left\{
\begin{array}{ll}
\displaystyle
J+J' =  \mu\sqrt{{\cal G}Ma(1-e^2)} + \mu'\sqrt{{\cal G}M'a'(1-e'^2)} & = {\rm constant} \\ \\
\displaystyle
\frac{J}{m} + \frac{J'}{m+1} \approx 
\frac{2}{mn_0} \cdot \left[-\frac{{\cal G}M\mu}{2a} - \frac{{\cal G}M'\mu'}{2a'} \right] & = {\rm constant.} \\
\end{array}
\right.
\label{eq_cons}
\end{equation}
We remind that this approximation is valid only when $\xi$, $\xi'$ $\ll 1/m$.
Consequently, the conservations of $J$ and $J'$ merely express 
the conservation of the total angular momentum and energy of the system.
More precisely, the Hamiltonian ${\cal H}_1$ describes the motion of two satellites that would orbit
a motionless central massive planet. Thus, the exchange of energy and angular momentum occurs
only between the satellites, and not between the satellites and the planet.
In terms of $\xi$, $e$, $\xi'$ and $e'$, the two equations (\ref{eq_cons}), under the assumption that $M \sim M'$, read:
\begin{equation}
\left\{
\begin{array}{ll}
\displaystyle
\frac{\mu(\xi -e^2)}{ a^2_0 n_0} + \frac{\mu'(\xi' -e'^2)}{ a'^2_0 n'_0} &= {\rm constant} \\ \\
\displaystyle
\frac{\mu \xi}{a_0} + \frac{\mu' \xi'}{a'_0} & = {\rm constant} \\
\end{array}
\right.
\label{eq_cons_xi_e}
\end{equation}

We now define 
$J_c = \Lambda + m\Gamma = J - m\Theta'$ and
$J'_c = \Lambda' - (m+1)\Gamma' = J' + (m+1)\Theta$,  
which are averaged versions of the Jacobi 
quantity, or Tisserand parameter, see e.g. \textcolor{blue}{\cite{mur00}}. 
It can be shown that, to within additive 
constants\footnote{The notation $J_c$ is used later for a local version of the 
Jacobi constant, see Table~(\ref{tab_eps_cn0}).
It is the same quantity as used here, except for a multiplicative constant.}:
\begin{equation}
\left\{
\begin{array}{ll}
\displaystyle
J_c = & 
\displaystyle
\frac{\mu a^2_0 n_0}{2} \left[\xi + m e^2\right] \\ \\
\displaystyle
J'_c = & 
\displaystyle
\frac{\mu' a'^2_0 n'_0}{2} \left[\xi' - (m+1) e'^2\right].
\end{array}
\right.
\label{eq_jac}
\end{equation}
The conservations of $J$ and $J'$ thus impose
$\dot{J}_c = m \partial {\cal H}_1/\partial \phi'$ and 
$\dot{J}'_c = -(m+1) \partial {\cal H}_1/\partial \phi$, or:
\begin{equation}
\left\{
\begin{array}{ll}
\displaystyle
\frac{d}{dt}\left(\xi + m e^2\right)= & 
\displaystyle
-\frac{2m  {\cal G}\mu'}{a^2_0 n_0} A' e' \cdot \sin(\phi') \\ \\
\displaystyle
\frac{d}{dt}\left[\xi' - (m+1) e'^2\right]=  &
\displaystyle
\frac{2(m+1)  {\cal G}\mu}{a'^2_0 n'_0}  A e \cdot \sin(\phi). \\
\end{array}
\right.
\label{eq_varia_jac}
\end{equation}
As expected, the Jacobi quantity $J_c$ (resp. $J'_c$) is constant if $e'=0$ (resp. $e=0$).

On the other hand, we have
$\dot{\Gamma} = \partial {\cal H}_1/\partial \varpi$ and
$\dot{\Gamma}' = \partial {\cal H}_1/\partial \varpi'$, so that:
\begin{equation}
\left\{
\begin{array}{ll}
\displaystyle
\frac{d(e^2\sqrt{{\cal G}Ma})}{dt}= &
2 {\cal G} \mu' A  e \cdot \sin(\phi) \\ \\
\displaystyle
\frac{d(e'^2\sqrt{{\cal G}M'a'})}{dt}= &
2 {\cal G} \mu A'  e' \cdot \sin(\phi'),
\end{array}
\right.
\label{eq_varia_jr}
\end{equation}

The quantity $e^2\sqrt{{\cal G}Ma}$ (resp.  $e'^2\sqrt{{\cal G}M'a'}$)
is the action associated with the fast, radial motions of $\mu$ (resp. $\mu'$).
In effect, the particle radial motion has an ampitude $\Delta r \approx ae$,
while its radial velocity has an amplitude $\Delta p_r \approx aen \propto a^{-1/2}e$,
from Kepler's third law. Thus, $\oint p_r dr \propto e^2a^{1/2}$.

In the case of a Keplerian central potential $-{\cal G}M_{p}/r$,
we have $\dot{\varpi}_s= \dot{\varpi}'_s=0$. 
The two-degree of freedom system described by ${\cal H}_1$
then admits a second integral of motion (besides the Hamiltonian itself), 
and is thus integrable.
This second integral was found by 
\textcolor{blue}{\cite{ses84}} for the general case $\mu \neq 0$, $\mu' \neq 0$
and extended to the restricted case by \textcolor{blue}{\cite{wis86}},
while being further analyzed by \textcolor{blue}{\cite{hen86}}.
The existence of this second constant can be demonstrated by using 
canonical transformations in which the sum
$\mu \mu'  A e \cdot \cos(\phi) + \mu \mu'  A' e' \cdot \cos(\phi')$ 
in Eq.~(\ref{eq_H1}) 
is replaced by a unique term $\sqrt{2\Phi} \cdot \cos(\varphi)$, 
and by showing that the new Hamiltonian only depends on the conjugate
variables $\Phi$, $\varphi$, reducing the system to a one-degree of freedom, 
integrable problem.

A more geometrical demonstration of the existence of a second constant of motion
for ${\cal H}_1$ is provided here by posing 
$\sigma= (m+1)\lambda' - m\lambda$ 
and by defining the vectors 
$\textbf{u}= [\cos(\sigma),\sin(\sigma)]$ and
$\textbf{v}= [-\sin(\sigma),\cos(\sigma)]$.
Note that $\partial \textbf{u}/\partial \sigma = -\textbf{v}$.
We also define the eccentricity vectors of $\mu$ and $\mu'$ as
\begin{equation}
\left\{
\begin{array}{ll}
\textbf{e}= &  (p,q)= [e \cos(\varpi), e  \sin(\varpi)]  \\ \\
\textbf{e}'=  & (p',q')= [e'  \cos(\varpi'), e'  \sin(\varpi')].
\end{array}
\right.
\label{eq_ecc_vec}
\end{equation}
For the Keplerian case, $\dot{\varpi}_s= \dot{\varpi}'_s=0$, the Hamiltonian ${\cal H}_1$  now reads:
\begin{equation}
{\cal H}_{1K}= 
-\frac{\mu^3({\cal G}M)^2}{2\Lambda^2} 
-\frac{\mu'^3({\cal G}M')^2}{2\Lambda'^2} + 
{\cal G} \mu \mu'  A   (\textbf{e} \cdot \textbf{u}) +
{\cal G} \mu \mu'  A'  (\textbf{e}' \cdot \textbf{u}).
\label{eq_H1K}
\end{equation}

Using the approximations $e \ll 1$ and 
$\Lambda \approx \Lambda_0 = \mu \sqrt{{\cal G}Ma_0}$, 
we obtain
$\dot{p}=  (\partial {\cal H}_{1K}/\partial q)/\Lambda_0$ and 
$\dot{q}= -(\partial {\cal H}_{1K}/\partial p)/\Lambda_0$, 
so that 
$\dot{\textbf{e}}= -({\cal G} \mu \mu' A/\Lambda_0) \textbf{v}$.
Likewise,  
$\dot{\textbf{e}'}= -({\cal G} \mu \mu' A'/\Lambda'_0) \textbf{v}$.
Using Kepler's third law and noting that $A$ and $A'$ have opposite signs, 
the two latter equations provides the following vectorial constant of motion:
\begin{equation}
\frac{\Lambda_0}{A} \textbf{e} - \frac{\Lambda'_0}{A'} \textbf{e}' \propto
\frac{\mu}{a_0 |A| n_0}\textbf{e} + \frac{\mu'}{a'_0 |A'| n'_0}\textbf{e}' = 
\textbf{e}_{\rm tot} = \rm constant,
\label{eq_sessin}
\end{equation}
where the vector $\textbf{e}_{\rm tot}$ defined here is called the ``total eccentricity" of the system.
Moreover, 
$\dot{\Lambda}= 
-\partial {\cal H}_{1K}/\partial \lambda = 
m \partial {\cal H}_{1K}/\partial \sigma =
- m(\Lambda_0 \textbf{e} \cdot \dot{\textbf{e}}  + \Lambda'_0  \textbf{e}' \cdot \dot{\textbf{e}'}).$
Using the conservation of $\textbf{e}_{\rm tot}$, and performing the same calculation for $\dot{\Lambda}'$, 
we obtain:
$$
\begin{array}{ll}
\displaystyle
\frac{A^2 \dot{\Lambda}}{m \Lambda_0}=           & -\left[A \textbf{e}  + A' \textbf{e}'\right]  \cdot A \dot{\textbf{e}} \\ \\
\displaystyle
\frac{A'^2 \dot{\Lambda}'}{(m+1) \Lambda'_0}=  & +\left[A \textbf{e}  + A' \textbf{e}'\right]  \cdot A' \dot{\textbf{e}'}.
\end{array}
$$
Subtracting these two equations and noting that 
$\dot{\Lambda}/\Lambda_0  \approx \dot{\xi}/2$ and  
$\dot{\Lambda}'/\Lambda'_0 \approx \dot{\xi}'/2$, we finally arrive at:
\begin{equation}
J_{c,\rm relat}=
\frac{A^2 \xi}{m} - \frac{A'^2 \xi'}{m+1} + \left(|A| \textbf{e}  - |A'| \textbf{e}'\right)^2 = \rm constant.
\label{eq_jac_relat_gen}
\end{equation}

Comparison with Eqs.~(\ref{eq_jac}) shows that the second constant of motion can be interpreted
as a ``relative Jacobi constant", which extends the notion of Jacobi constant to the non-circular 
3-body problem.
Eq.~(\ref{eq_jac_relat_gen}) tells us that the exchange of energy between the satellites 
only depends on the relative eccentricity vector $\textbf{e}_{\rm relat}= |A| \textbf{e}  - |A'| \textbf{e}'$.
Moreover, Eq.~(\ref{eq_cons_xi_e})  tells us how this energy is distributed between $\mu$ and $\mu'$.
Finally, Eq.~(\ref{eq_sessin}) shows that the interaction between $\mu$ and $\mu'$
conserves the total eccentricity of the system $\textbf{e}_{\rm tot}$.
For large $m$'s, $|a_0 A n_0| \approx |a'_0 A' n'_0|$, and the total eccentricity 
is just proportional to $\mu \textbf{e} + \mu' \textbf{e}'$.
This is a classical result already obtained by \textcolor{blue}{\cite{heno86}} for the Hill's Keplerian limiting case 
and by \textcolor{blue}{\cite{for96}} for the generalization to an oblate planet.


\section{Restricted case}

When $\mu = 0$,
the actions describing the motion of $\mu$ must be
expressed in terms of unit mass,
we remind that $\mu'$ is a dimensionless parameter. 
Thus, the hamiltonian ${\cal H}_1$
must be divided by $\mu$. 
This new hamiltonian is not autonomous anymore, 
since $\lambda'$ and $\varpi'$ are now linear functions of time: 
$\lambda'= n' t$ and $\varpi'= \dot{\varpi}'_s t$.
This yields the Hamiltonian:
\begin{equation}
{\cal H}_2= 
-\frac{({\cal G}M)^2}{2\Lambda^2}
-\frac{\mu'({\cal G}M')^2}{2\mu\Lambda'^2} +
{\cal G} \mu'  A \sqrt{\frac{2\Gamma}{\Lambda}} \cdot \cos(\phi) + 
{\cal G} \mu'  A' e' \cdot \cos(\phi')
- \dot{\varpi}_s\Gamma 
+ \frac{n' \Lambda'}{\mu} - \dot{\varpi}'_s\Gamma',
\label{eq_H2}
\end{equation}
where the angle-action variables are now:
\begin{equation}
\begin{array}{rl}
\lambda  \longleftrightarrow & \Lambda = \sqrt{{\cal G}Ma} \\
-\varpi     \longleftrightarrow & \Gamma= \sqrt{{\cal G}Ma}(1 -\sqrt{1-e^2}) \approx  e^2\sqrt{{\cal G}Ma}/2 \\
\lambda'= n't   \longleftrightarrow & \Lambda' =  \sqrt{{\cal G}M'a'}\\
-\varpi' =  -\dot{\varpi}'_s t   \longleftrightarrow & \Gamma'  =  \sqrt{{\cal G}M'a'}(1 -\sqrt{1-e'^2}) \approx  e'^2\sqrt{{\cal G}M'a'}/2 \\
\label{eq_Poin_res}
\end{array}
\end{equation}

Note that for $\mu = 0$, $\Lambda'$ and $\Gamma'$ being constant, the corresponding terms above can be dropped from the Hamiltonian.
Then, the treatment of ${\cal H}_2$ proceeds in the same way as for the general case
previously considered, 
i.e. by defining the same transformations as in Eqs.~(\ref{eq_newvar}).
Again the Hamiltonian is reduced to that of a two-degree of freedom system which is 
in general not integrable, except for the special case of a central Keplerian potential $-{\cal G}M_{p}/r$.
In this case, $\xi'$ and $\textbf{e}'$ are constant, so that Eq.~(\ref{eq_jac_relat_gen}) can be re-written:
\begin{equation}
J_{c,\rm relat}=
\xi + m \left(\textbf{e} - \left|\frac{A'}{A}\right|\textbf{e}'\right)^2 = \rm constant,
\label{eq_jac_relat_res}
\end{equation}
which generalizes the expression of the Jacobi constant $\xi + me^2$ associated with $\mu$, 
where the eccentricity vector $\textbf{e}$ has been replaced by the
relative eccentricity vector 
\begin{equation}
\textbf{e}_{\rm relat} = \textbf{e} - \left|\frac{A'}{A}\right|\textbf{e}'.
\label{eq_e_relat}
\end{equation}
Physically, this means that, at the first order approximation in eccentricities used here,  
the exchange of energy between the satellite and the particle (described by $\dot{\xi}$)
only depends on the relative eccentricity vector $\textbf{e}_{\rm rel}$.


Finally, both Eqs.~(\ref{eq_jac_relat_gen}) and (\ref{eq_jac_relat_res}) show why the relative 
Jacobi constant is destroyed when the central potential departs from the Keplerian form $-{\cal G}M_{p}/r$.
In fact, a general potential induces a differential secular precession rate $\dot{\varpi}_s - \dot{\varpi}'_s$
between the vectors $\textbf{e}$ and $\textbf{e}'$, which causes a drift of the angle between
$\textbf{e}$ and $\textbf{e}'$ which is imposed ``from outside", 
i.e. independent of the interactions between $\mu$ and $\mu'$.

We conjecture that as soon as $\dot{\varpi}_s \neq \dot{\varpi}'_s$, 
the Hamiltonians ${\cal H}_1$ and ${\cal H}_2$ (Eqs.~(\ref{eq_H1}) and (\ref{eq_H2})) 
are not integrable, as shown for instance in Fig.~(\ref{sum}).
The demonstration of such a result is beyond the scope of this paper, however,
and will be accepted on numerical grounds.


\section{Lindblad vs. corotation resonances}
\label{sec_lin_vs_cor}

Eqs.~(\ref{eq_varia_jac}) and (\ref{eq_varia_jr}) show that the term
$A' \cdot \sin{(\phi'})$ 
modifies the Jacobi quantity of $\mu$ and  the radial action of $\mu'$.
Conversely, the term 
$A \cdot \sin{(\phi})$ 
modifies the Jacobi quantity of $\mu'$ and  the radial action of $\mu$.
Consequently, the resonances associated with $\phi$ and $\phi'$ play symmetrical roles, 
and cannot be distinguished, as it is clearly apparent from the form of ${\cal H}_1$  (Eq.~(\ref{eq_H1})).

This symmetry is broken, however,  when the mass of one of the satellites tends to zero 
(restricted case), and the different roles played by the two resonances
clearly appear.
Taking for instance $\mu=0$,
the resonance associated with $\phi$ is then called the Lindblad Eccentric Resonance (LER), while
the resonance associated with $\phi$ is called the Corotation Eccentric Resonance 
(CER)\footnote{Lindblad and corotation resonances associated with orbital inclinations are also possible,
hence the specific term ``eccentric" used here.}.
This nomenclature comes from galactic and ring dynamics, 
where those resonances were studied, see \textcolor{blue}{\cite{gol79}} and
the references already quoted in the introduction.

If we consider the LER alone (i.e. taking $A'=0$), then 
$\xi + me^2$ is constant. Using $\xi = (a-a_0)/a_0$, this yields
\begin{equation}
\frac{\delta a}{a} = -2m e^2 \cdot \frac{\delta e}{e}.
\label{eq_eLin}
\end{equation}
Thus, the LER  mainly excites the orbital eccentricity of $\mu$, 
and much less its semi-major axis.
This is physically understandable by noting that the LER  corresponds to
$(m+1)n' -mn - \dot{\varpi}_s = 0$, i.e.  $\kappa= (m+1)(n-n')$, where
$\kappa=  n - \dot{\varpi}_s$ is the epicyclic, radial oscillation of the particle.
The quantity $n-n'$ is the synodic frequency, i.e.  the frequency at which the
satellite and the particle are in conjunction. 
Thus, $n-n'$ is the frequency at which $\mu'$ perturbs $\mu$ through periodic kicks 
in the \textit{radial} direction.
For $\kappa= (m+1)(n-n')$, those kicks resonantly excite the radial action of $\mu$ (Eq.~(\ref{eq_varia_jr})).
On the other hand, because they are radial, the kicks essentially conserve the energy of the particle, 
and thus, its semi-major axis, as shown in Eq.~(\ref{eq_eLin}).

Conversely, let us consider the case where the LER is far away from the CER. Then $\phi$ varies rapidly in ${\cal H}_2$, and the corresponding term can be zeroed, which is equivalent to taking $A=0$, 
the radial action $\Gamma \propto e^2 a^{1/2}$ is constant.
This imposes
\begin{equation}
\frac{\delta e}{e} =  -\frac{\delta a}{4a}.
\label{eq_ecor}
\end{equation}
The comparison of Eqs.~(\ref{eq_eLin}) and (\ref{eq_ecor}) shows that the corotation 
resonance affects much less the orbital eccentricity of the particle than the Lindblad resonance.

This is physically understandable by noting that the CER corresponds to
$(m+1)n' -mn - \dot{\varpi}'_s = 0$, i.e.  $n= n' + \kappa'/m$, where
$\kappa'=  n' - \dot{\varpi}'_s$ is the epicyclic, radial oscillation of the satellite $\mu'$.
Then, the mean motion of $\mu$ matches the pattern speed $n_{\rm pattern} = n' + \kappa'/m$ 
of one of the harmonics of the disturbing potential of $\mu'$, hence the denomination ÒcorotationÓ.
Those resonances are in fact identical in essence to the classical 1:1 co-orbital resonance, 
but they occur at radii that  are different from $a'$.

Note in particular that the CER slowly modulates the \textit{azimuthal} 
potential acting on the particle since
${\cal G}\mu \mu' e' A' \cdot \cos(\phi') \approx {\cal G}\mu \mu' e' A' \cdot \cos[m(\theta- n_{\rm pattern} t)]$,
where $\theta$ is the true longitude of $\mu$.
This periodic potential creates a small, slowly varying azimuthal acceleration on the particle
that slowly modifies its semi-major axis $a$.
As $a$ slowly varies, the radial action of $\mu$, $\Gamma \propto e^2 a^{1/2}$ is conserved,
as expressed in Eq.~(\ref{eq_ecor}).
More exactly, this conservation is actually the adiabatic conservation of the fast radial action
$e^2 a^{1/2}$, as the azimuthal action $\sqrt{{\cal G}M_{p}a}$ slowly varies. 
More discussions about the conservation of $e^2a^{1/2}$ in various contexts
can be found in \textcolor{blue}{\cite{fle00}} and  \textcolor{blue}{\cite{sic03}}.

While the separate effects of LER and CER are easy to describe
in term of 1-degree of freedom systems, the problem is complex when they are coupled. 
The following section provides the simplest  equations that permit to explore this complexity.


\section{The CoraLin model}
\label{sec_coralin}
 
The restricted case $\mu = 0$ can be studied through the Hamiltonian 
${\cal H}_2$  (Eq.~(\ref{eq_H2})).
The derivation of the equations of motion stemming from ${\cal H}_2$
 is standard, see for instance \textcolor{blue}{\cite{mur00}}.
Near $a_0$, the actions $\Lambda$ and $J$ can be written   
$\Lambda =  \Lambda_0 + \Delta \Lambda$ and
$J= J_0 + \Delta J$, respectively, where 
$\Delta J=  \Delta \Lambda + m(\Theta + \Theta')$ is a constant of motion.
The Hamiltonian ${\cal H}_2$ is then expanded to second order in $\Delta \Lambda$, 
providing an approximation of the Hamiltonian valid near $a_0$.

At this point, it is useful to consider the Jacobi quantity $J_c= \Delta \Lambda + m\Gamma$ 
that appears in the first line of Eqs.~(\ref{eq_newvar}),
from which we obtain $J_c=  \Delta J - m\Theta'$.
Consequently, 
$\dot{J_c}=  -m\dot{\Theta}'= m \partial {\cal H}_2/\partial \phi'$ and
$\dot{\phi}' = -m\partial {\cal H}_2/\partial J_c$.
These are almost the canonical Hamiltonian equations, but not quite, 
because of the appearance of the factor $-m$.
This suggests to take $\phi'$ and $J_c$ as conjugate variables, 
after redefining the action $J_c$ and the Hamiltonian ${\cal H}_2$
to within multiplicative and additive factors.
The choice of those factors is rather arbitrary. 
We choose them so that to simplify as much as possible the expression of ${\cal H}_2$,
so that to obtain the form in Eq.~(\ref{eq_Hnorm}).
Moreover, as the particle remains close to the corotation resonance radius $a_0$, 
it is convenient to use the new time scale $\tau= n_c t$, where $n_c$ is
the libration frequency of $\phi_c$ near the corotation fixed point 
in the absence of the Lindblad resonance, see Table~\ref{tab_eps_cn0}.
 
As discussed in Section~\ref{sec_lin_vs_cor}, 
the resonances associated with $\phi$ and $\phi'$
can be separated  into LER and CER types, respectively.
To enhance this distinction, we will use from now the conjugate variables 
($\phi_c$,$\phi_L$,$J_c$,$J_L$) 
instead of ($\phi^{'}$,$\phi$,$\Theta^{'}$,$\Theta$).
The actions $J_c$ (proportional to $\Delta \Lambda + m\Gamma)$ and 
$J_L$ (proportional to $\Theta$) are defined in Table~\ref{tab_eps_cn0}.
Moreover, the angles $\phi_c$ and $\phi_L$ are defined by:
\begin{equation}
\left\{
\begin{tabular}{llll}
$\phi_c=$ & $+\phi' + \pi$  & $= +(m+1)\lambda' - m\lambda - \varpi'+\pi$ & if  $m>0$ ($\mu$ inside $\mu'$) \\ 
$\phi_c=$ & $+\phi'         $ & $= +(m+1)\lambda' - m\lambda - \varpi'$       &  if  $m<0$ ($\mu$ outside $\mu'$) \\ 
$\phi_L=$ & $-\phi          $  & $=  -(m+1)\lambda' + m\lambda + \varpi$   &  \\ 
\end{tabular}
\right.
\end{equation}

The particular choice for $\phi_c$ is motivated by the fact that it allows a unique
form of ${\cal H}$, avoiding a $\pm 1$ factor in front of
the term $\cos(\phi_c)$ in Eq.~(\ref{eq_Hnorm}).
With this convention, the stable corotation point is always at $\phi_c=0$.
Moreover, the minus sign used to define $\phi_L$ from $\phi$ stems from the requirement 
that  we retrieve the canonical equations  
$\dot{h} = -\partial {\cal H}/\partial k$ and
$\dot{k} = +\partial {\cal H}/\partial h$
with the correct signs in the system~(\ref{eq_coralin0}).
%
\begin{table}[h!]
\centering
\renewcommand{\arraystretch}{2.1}		
\begin{tabular*}{140mm}{ccc}
\hline
\multicolumn{3}{c} {Constant parameters} \\
\hline
$\displaystyle n_c= \left(3m^2  a_0 |A'| e' \frac{\mu'}{M}\right)^{1/2} \cdot n_0$ & 
$\displaystyle D= \frac{(\dot{\varpi}'_s -\dot{\varpi}_s)}{ n_{c}}  $  & 
$\displaystyle \epsilon_L= 
\left(\frac{a_0}{3m^2} \frac{\mu'}{M}\right)^{1/4} \cdot
\frac{A}{\left(|A'|e'\right)^{3/4}}$ \\ [2mm]
\hline
\multicolumn{3}{c} {Actions} \\
\hline
\multicolumn{2}{r}{$\displaystyle J_c= 
{\rm sgn}(m) \left(\frac{3}{4a_0|A'|}\frac{M}{\mu' e'}\right)^{1/2} \cdot (\xi + me^2)$} & 
$J_L= \left(\frac{3m^2}{2a_0|A'|} \frac{M}{\mu' e'}\right)^{1/2}  \cdot e^{2}$ \\ [2mm]
\hline
\end{tabular*}
\caption{
Quantities entering in the Hamiltonian ${\cal H}$ (Eq.~\ref{eq_Hnorm}),
where $A$ and $A'$ are a combination of Laplace coefficients given by (\ref{eq_AA'}) and $a_{0}$ is the reference value of the corotation,
Note that the time scale used in Eq.~(\ref{eq_coralin0}) is $\tau= n_c t$.
Those definitions relate the actual parameters of the problems (mass of the satellite,
semi-major axis, orbital eccentricity, etc..) and the non-dimentional variables used in 
Eqs.~(\ref{eq_Hnorm}) and (\ref{eq_coralin0}).
}
\label{tab_eps_cn0}
\end{table}

\begin{figure}[!h]
\centerline{\includegraphics[angle=0,totalheight=7cm,trim= 0 0 0 0]{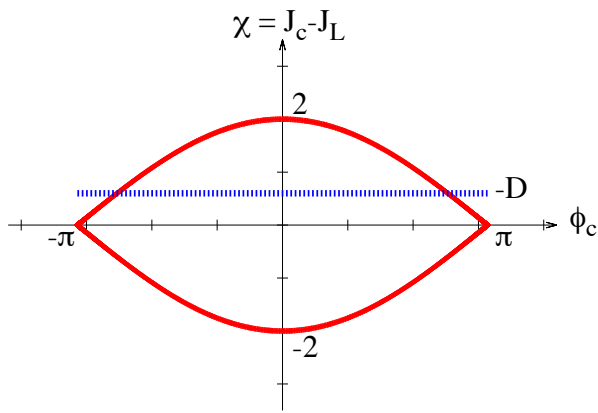}} 
\caption{%
Scheme of the CoraLin model.
In the absence of the Lindblad resonance (LER), the pendular
motion forced by the corotation resonance (CER) is confined into the red separatrix curve, 
whose full width is given by  $\Delta \chi = \pm 2$.
In the absence of the CER, the LER radius is at  $\chi= -D$ (dotted blue line)
The coupling between the two resonance strongly depends on $D$, see Fig.~(\ref{sum}). 
}%
\label{fig_coralin}
\end{figure}

Using the equations
$(m+1)n'_0 -mn_0 - \dot{\varpi}'_s=0$ and 
$(m+1)n'_0 -mn_0 - \dot{\varpi}_s= \dot{\varpi}'_s - \dot{\varpi}_s$ (see Section~(\ref{sec_hal})), and
the quantities $D$, $\epsilon_L$, $J_c$ and $J_L$ defined in Table~\ref{tab_eps_cn0}, 
we finally obtain the following Hamiltonian with the two pairs of conjugate variables
$(J_c,\phi_{c})$ and 
$[h=\sqrt{2 J_L} \cdot \cos(\phi_L),k=\sqrt{2 J_L} \cdot \sin(\phi_L)]$:
\begin{equation}
{\cal H}= 
\frac{(J_c - J_L)^2}{2} 
- D J_L
-  \cos(\phi_{c}) 
- \epsilon_L h.
\label{eq_Hnorm}
\end{equation}
The associated equations of motion are:
\begin{equation}
\left\{
\begin{array}{lll}
\displaystyle \frac{dJ_c}{d\tau}=  & \displaystyle -\frac{\partial {\cal H}}{\partial \phi_c}=  &  -\sin(\phi_c) \\
& \\
\displaystyle
\frac{d\phi_{c}}{d\tau}= & \displaystyle +\frac{\partial {\cal H}}{\partial J_c}=  & J_c - [J_L] \\
& \\
\displaystyle
\frac{dh}{d\tau} = & \displaystyle -\frac{\partial {\cal H}}{\partial k}= & +([J_c] - J_L + D)  k \\
& \\
\displaystyle
\frac{dk}{d\tau} = & \displaystyle +\frac{\partial {\cal H}}{\partial h}= & -([J_c] - J_L + D) h  - \epsilon_L, \\
\end{array}
\right.
\label{eq_coralin0}
\end{equation}
where $\tau= n_c t$ and $J_L= (h^2 + k^2)/2$.
Note from Eqs.~(\ref{eq_AA'}) and Table~\ref{tab_eps_cn0} that ${\rm sgn}(\epsilon_L) = {\rm sgn}(m)$.
Note that $D$ (defined in table \ref{tab_eps_cn0}) depends on both $e'$ and $\mu'$. Our choice of $D$ is such that the width of corotation resonance is fixed to $\pm2$, (see fig. \ref{fig_coralin}).

We call this system of equations the ``CoraLin" model, as it describes the motion of a particle
near a corotation and a Lindblad resonances, 
for which, the centers occur respectively at $J_{c} = J_{L}$ and $J_{c}-J_{L} = -D$.
It is completely non-dimensional and can be used in a generic way to analyze 
the coupling between the two resonances. 
In fact, this system is parametrized by only two quantities: 
$D$, which measures the distance between the two resonances, and
$\epsilon_L$, which measures the forcing of the particle orbital eccentricity by the satellite,
while absorbing the satellite's  orbital eccentricity.
Finally, the time scale $\tau$ is parametrized by the quantity $n_c$, see Table~\ref{tab_eps_cn0}.

The coupling between the two resonances comes from the bracketed terms in Eqs.~(\ref{eq_coralin0}), 
namely
$(i)$ the term $J_L$ in the second equation, 
which tells us how the particle orbital eccentricity (mainly driven by the Lindblad resonance) 
perturbs the simple pendulum motion and
$(ii)$ the term $J_c$ in the third and fourth equations, 
which tells us how the corotation resonance affects the motion of $(h,k)$ driven by 
the Lindblad resonance. 

If we suppress the term $\cos (\phi_c)$ in ${\cal H}$ in order to keep only the LER, 
then the Hamiltonian takes the form
\begin{equation}
{\cal H}_{\rm And}= J_L^2/2 - (J_c + D)J_L - \epsilon_L h,
\label{eq_H_And}
\end{equation}
where $J_c$ is now a constant parameter.    
This is the classical Andoyer Hamiltonian that has been extensively 
studied and reviewed in many publications in the last few decades,
see e.g. \textcolor{blue}{\cite{hen83}}, \textcolor{blue}{\cite{fer85}} and \textcolor{blue}{\cite{fer07}}.

On the other hand, if we make $\epsilon_L=0$, then $J_L$ is constant,
${\cal H}$ reduces to the Hamiltonian of the simpe pendulum:
\begin{equation}
{\cal H}_{\rm pen}=  \chi^2/2 -   \cos(\phi_{c}),
\label{eq_H_pend}
\end{equation}
where we define $\chi= J_c- J_L$.
Then we obtain $\ddot{\phi}_c= -  \sin(\phi_c)$, which describes the stable oscillatory motion 
of the particle guiding center around the corotation fixed point at $\phi_c=0$.
Note that in this case, $J_L=$ constant, meaning that the particle orbital eccentricity is conserved, 
as  announced by Eq.~(\ref{eq_ecor}).
 
An alternative form of the system~(\ref{eq_coralin0}), although not using conjugate variables, is:
\begin{equation}
\left\{
\begin{array}{ll}
\dot{\chi}=  &  -\sin(\phi_{c}) - \dot{J}_L\\
& \\
\dot{\phi}_{c}= & \chi \\
& \\
\dot{h} = & -(\chi + D)  k \\
& \\
\dot{k} = & +(\chi+ D) h  + \epsilon_L. \\
\end{array}
\right.
\label{eq_allchi}
\end{equation}

If the corotation motion of the particle is not perturbed by the LER,
then the librating zone for  $\phi_c$ has a width of $\Delta \chi = \pm 2$,
see Fig.~(\ref{fig_coralin}).
The exact LER occurs at $\chi = -D$.
Consequently, the two resonances collapse into a single one for $D=0$, and
are well separated for $|D|$ significantly larger than 2.
We now explore the dynamics of the system for intermediate values of $D$, showing 
that significant chaotic zones appear in the phase space for those intermediate cases.


\section{Asymptotic behaviors}
\label{sec_asym}

 
\subsection{Superimposed resonances}

If $D=0$, the CER and LER are superimposed, and the problem is integrable (Section~\ref{sec_const}). 
Poincar\'e surfaces of section taken at $k=0$ are then regular, and they can be obtained analytically by noting that the
second constant of motion in Eq.~(\ref {eq_jac_relat_res}) can be re-written:
\begin{equation}
S = J_{c} + \frac{\sqrt{2J_{L}}}{\epsilon_{L}}\cos(\phi_{c}-\phi_{L}),
\label{eq_Sessin}
\end{equation}
which is another expression of the relative Jacobi constant defined in equation (\ref{eq_jac_relat_res}).

Using Eqs.~(\ref{eq_Hnorm}) and (\ref {eq_Sessin}) and making $k=0$, we obtain: 
\begin{equation}
\chi^4 - 4H\chi^2 + 8\epsilon_{L}^2\chi - 4\cos^2(\phi_{c}) + 4H^2 = 8\epsilon_{L}^2 S,
\label{H_S}
\end{equation}
where $H$ is the value of the Hamiltonian ${\cal H}$. 
For $H$ fixed, the surface of sections are the level contours of the surface defined by Eq.~(\ref{H_S})
for various values of $S$.
Note these contours are $\pi$-periodic, not $2\pi$-periodic.
The fixed points are given by the singular points of that surface: 
$$
\left\{
\begin{array}{ll}
\chi^3 - 2H\chi +2\epsilon_{L}^2 = 0   \\
& \\
 \phi_{c} = k\frac{\pi}{2}, ~~~~~~ k \in \mathbb{Z} \\
\end{array}
\right.
$$
The number of solutions in $\chi$ depends of the sign of the discriminant 
$\Delta = 32H^3 - 108\epsilon_{L}^4$.
Therefore, there exists a critical value $H_{0}=(27\epsilon_{L}^4/8)^{1/3}$:
for $H<H_{0}$ the system has only one solution in $\chi$:
$$
\chi_{0} = 
\left(\frac{2\epsilon_{L}^2 + \sqrt{-\Delta/27}}{2}\right)^{1/3} + 
\left(\frac{2\epsilon_{L}^2 - \sqrt{-\Delta/27}}{2}\right)^{1/3}
$$
For $H>H_{0}$ a pitchfork bifurcation occurs and provides three solutions:
$$
\chi_{p} = 
2\sqrt{\frac{-2H}{3}}
\cos\left[
\frac{1}{3}
\arccos
\left(-\epsilon_{L}^2\sqrt{27/8H^3}\right) + \frac{2p\pi}{3}
\right]
$$
with $p \in \{0,1,2\}$.

\begin{figure}[!h]
\centerline{\includegraphics[angle=0,totalheight=7cm,,width=15cm,trim= 0 0 0 0]{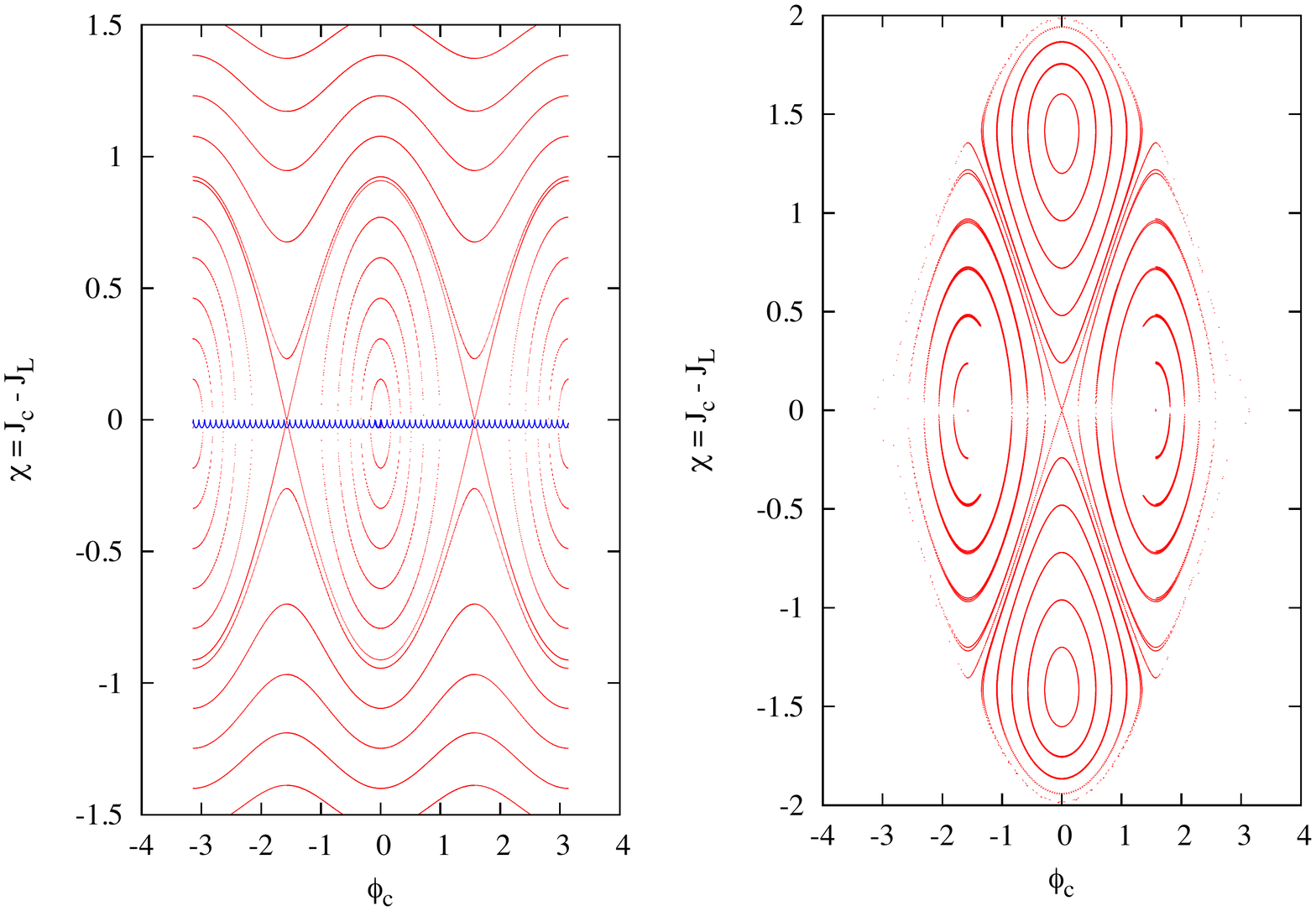}}
\caption{%
Phase portraits of the system~(\ref{eq_coralin0}) when the corotation and Lindblad
resonances (CER and LER, respectively) are superimposed, i.e. $D=0$.
We show Poincar\'e surfaces of section corresponding to $k=0$, 
with prescribed values $H$ of the Hamiltonian ${\cal H}$, and 
various values of the second constant of motion $S$ (Eq.~(\ref{eq_Sessin})).
For both panels,  $\epsilon_{L} = -0.123$, 
which provides a critical value $H_{0}=(27\epsilon_{L}^4/8)^{1/3}= 0.091...$.
\color{black} 
Left panel: $H = -1 < H_0$, there is only one solution in $\chi$ for the fixed points.
Right panel: $H=1 > H_0$, there are three solutions in $\chi$ for the fixed points.
See text for details.
}%
\label{D0}
\end{figure}

Examples of surfaces of section with $H<H_{0}$ and $H>H_{0}$ are given in Fig.~\ref{D0}. 
Note that orbits near the fixed elliptical points do \textit{not} correspond to librations of $\phi_c$
(i.e. to particles trapped in a CER).
For instance the fixed point near $\phi_c=0$, $\chi=0$ in the left panel corresponds to the trajectory
shown in blue, for which  $\phi_c$ is circulating.

\subsection{Well separated resonances}

We consider the situation in which the CER and LER are well separated, $|D| \gg 2$. 
Two cases are discussed.
 
\begin{figure}[!h]
\centerline{\includegraphics[angle=0,totalheight=7cm,width=14cm,trim= 0 0 0 0]{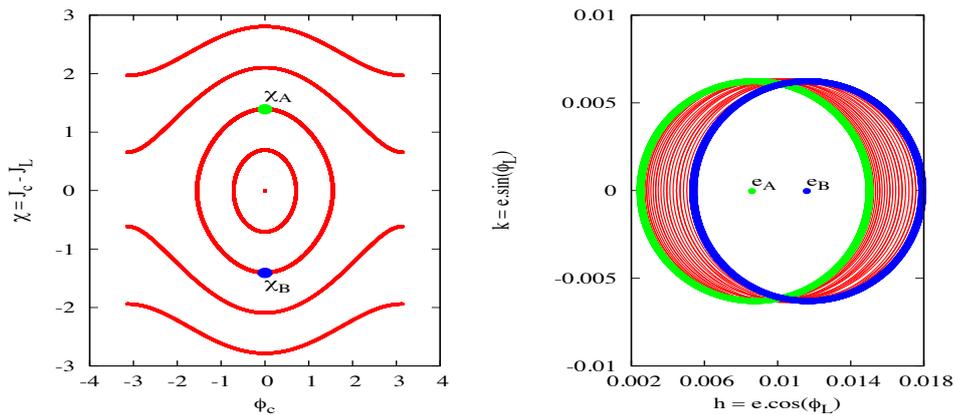}}    
\caption{%
Left:
Poincar\'e surfaces of section $k=0$ of the system~(\ref{eq_coralin0}) 
with $D = 10$ (the Lindblad resonance is far away from the corotation radius at $\chi=-D=-10$)
and $\epsilon_{L} = -0.1$. 
The phase portrait of the CER is very close to that of a simple pendulum. 
Right : 
the vector $(h,k)$ rapidly describes a circle centered on the forced value 
($h= e_f= |\epsilon_L/(\chi+D|,k=0$) that moves slowly as $\chi$ varies (red circles).
When 
$\chi = 1.4 $ (green point at left), then $e_{f} = e_{A} = (0.008,0)$, corresponding to the green orbit at right.
When 
$\chi = -1.4 $ (blue point at left), then $e_{f}  = e_{B} = (0.016,0)$ (blue orbit at right).
}%
\label{cer}
\end{figure}

(a) \textit{The particle is trapped in the corotation region.}
In this case, the variations of $(h,k)$ in the systems~(\ref{eq_coralin0}) and (\ref{eq_allchi}) 
are  much faster than the variations of $(J_c,\phi_{c})$. 
Consequently, the action $\oint hdk$ is adiabatically conserved.
Since the vector $(h,k)$  essentially describes a circle centered on the forced value 
$(-\epsilon_L/(\chi+D),0)$,
it means that $(h,k)$ rapidly moves on a circle of constant radius, 
whose  center slowly moves along the $Oh$ axis, see Fig.~\ref{cer}.
In particular, if $(h,k)$ starts at the forced value $(-\epsilon_L/(\chi+D),0)$,
then it will stay at that value as $\chi$ slowly changes.
In other words, the orbital eccentricity of the particle will permanently adjust itself so that 
$e= |\epsilon_L/(\chi+D|$ as $\chi$ varies. 
\begin{figure}[!h]
\centerline{\includegraphics[angle=0,totalheight=6cm,trim= 0 0 0 0]{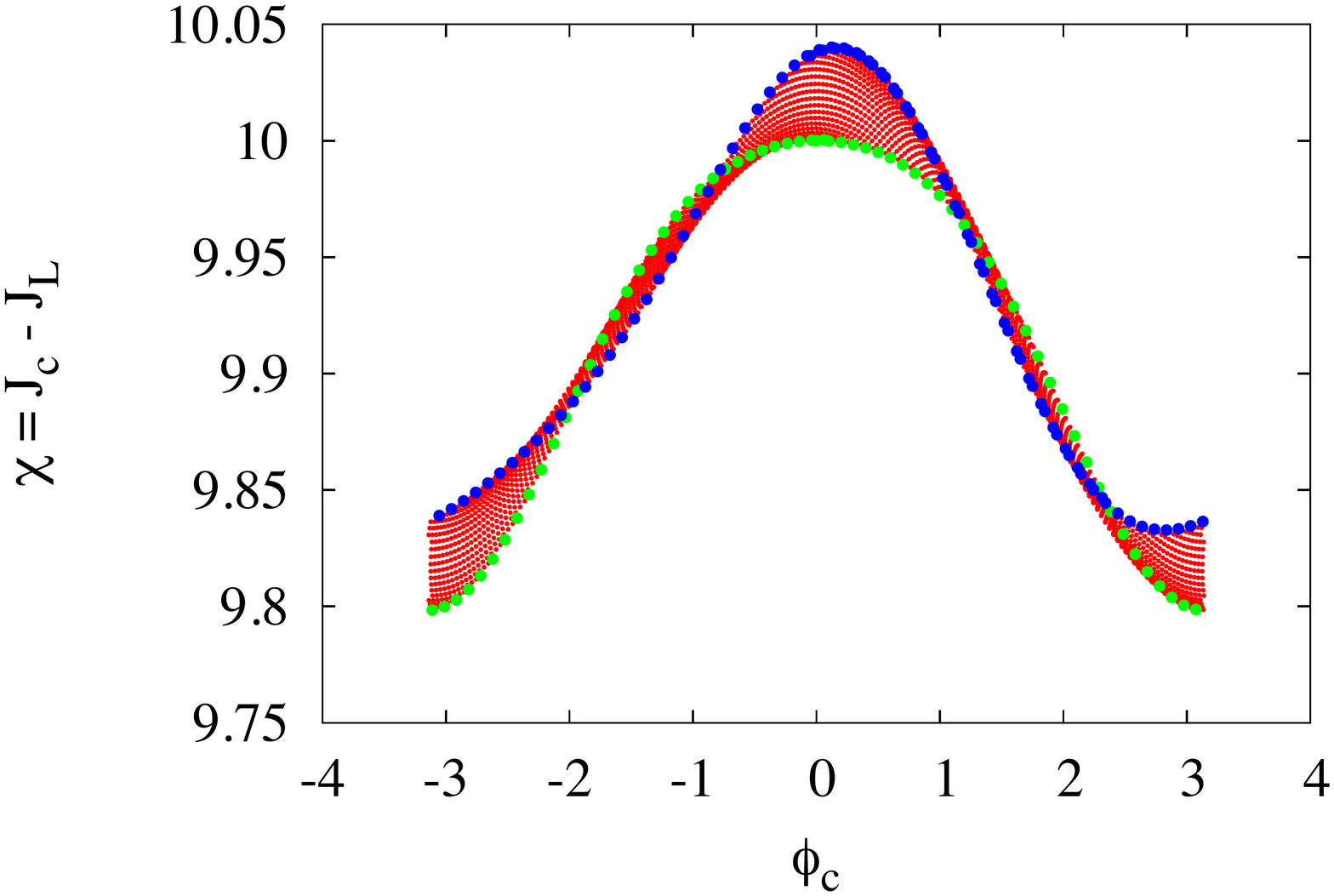}}
\caption{%
Poincar\'e surfaces of section $k=0$ of the system~(\ref{eq_coralin0}) 
with $D = 10$ and $\epsilon_{L} = -0.1$.
The particle is now close to the Lindblad resonance, $\phi_c$ and $\chi$ vary rapidly
compared to $h$ and $k$,  while keeping $\int J_c d\phi_{c}$ adiabatically constant.
}%
\label{ler}
\end{figure}

(b) \textit{The particle is trapped in the Lindblad resonance.}
The situation is now reversed: $(h,k)$ slowly varies as $J_c$ as $\phi_{c}$ oscillates rapidly. 
Thus $\int J_c d\phi_{c}$ is adiabatically conserved. In fact, the system : 
\begin{equation}
\left\{
\begin{array}{ll}
 \dot{J}_{c} =-\sin(\phi_{c})   \\ \\
 \dot{\phi}_{c} =  J_{c} - J_{L}
\end{array}
\right.
\label{eq_pend}
\end{equation} 
correspond to a simple pendulum with a slowly varying parameter, $J_L$.
Thus, the particle will librate around the slowly varying point  $(J_c= J_L, \phi_{c}=0)$, 
while adiabatically conserving $\int J_c d\phi_{c}$, see Fig.~\ref{ler}.
%
%


\section{Intermediate cases}
\label{sec_inter}

\subsection{Chaoticity}

We have explored numerically the transition from the integrable case $D=0$ to the chaotic regime for $D$ of order unity.
We consider here an illustrative case where $\epsilon_{L} = -0.1$ and $H = -0.5$. 
The value of $H$ using the orbital elements is :
\begin{center}
$\displaystyle
H =   \frac{(3m\Delta a/a_{0})^{2} }{8\sqrt{\epsilon_{c}}}  - \frac{3Dm^{2}e^{2}}{2\sqrt{\epsilon_{c}}} - \cos{\phi_{c}} - \epsilon_{L}\frac{\sqrt{3m^{2}e^{2}}}{{\epsilon_{c}}^{1/4}}\cos{\phi_{L}}$,
\end{center}
with $\epsilon_{c} = 3m^{2}a_{0}|A'| e' \mu'/M$.
This choice is motivated by the fact that they are typical values relevant to the small Saturnian satellites 
Anthe, Methone and Aegaeon, see Table~\ref{moons_param}.

\begin{figure}[!h]
\centerline{\includegraphics[angle=0,width= 8cm,trim= 0 0 0 0]{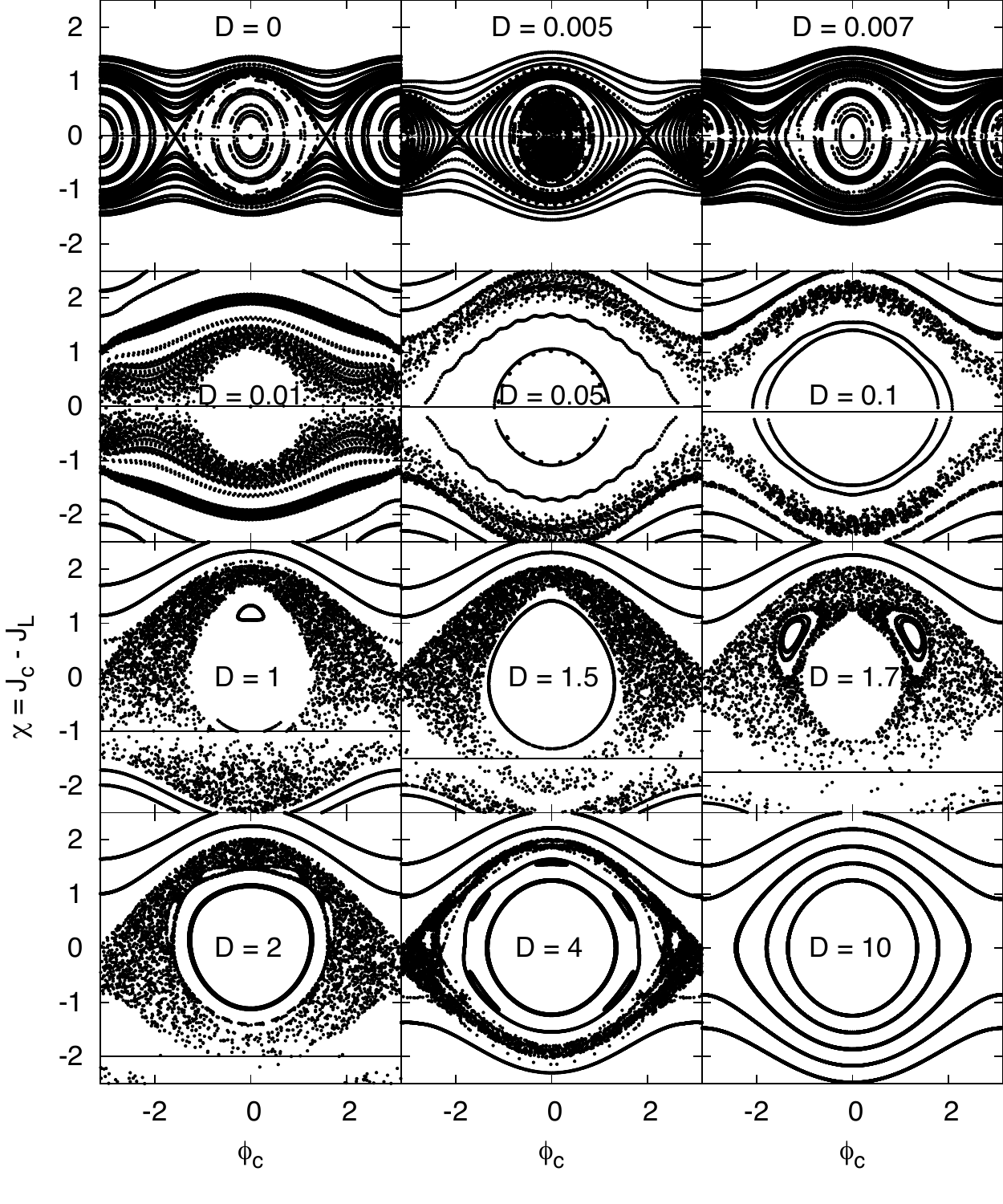}}    
\caption{%
Poincar\'e surfaces of section of system~(\ref{eq_coralin0}) using 
$\epsilon_{L} = -0.1$ and $H=-0.5$ 
for twelve different values of $D$.
Each phase portrait is obtained when $k=0$ for sixteen different trajectories with the same value of $H$.
When $D = 0$ and $D=10$, the trajectories are regular.
For intermediate cases, chaos is prevalent. In the CoraLin system (\ref{eq_coralin0}), we can see that for $D=0$, the vector $(h,k)$ moves slowly, so that $k$ passes more rarely through zero, explaining the rarefaction of points along the line $D=0$.
}%
\label{sum}
\end{figure}

Fig. ~\ref{sum} shows that the phase portrait of system~(\ref{eq_coralin0}) is rapidly invaded by a chaotic region 
for values of $D$ as small as $\sim 0.01$.
As $D$ increases, a central regular region appears, corresponding to the trapping of the particle in the corotation
site, i.e. to a libration of the critical angle $\phi_c$. 
Only for value of $|D|$ significantly larger than 2 does the system retrieves its regularity, 
and can the orbits be described using adiabatic invariant arguments, see Section~\ref{sec_asym}.
In fact, Fig.~(\ref{sum}) shows that for $0< |D| < 2$ the motion of the particle near the 
CER is dominated by chaos.
This is true for the actual Saturnian satellites that we examine now.

\subsection{Numerical exploration of the chaoticity}

In order to give a better and a global view of the dynamics in the intermediate cases, we perform a more systematic analysis of the dynamics for all of the regimes: the two limit cases (CER and LER) and the intermediate regime by measuring the Fast Lyapunov Indicator (FLI) detailed in (\textcolor{blue}{\cite{Froeschle97} and \cite{morby02}}).
In practice, we have study the FLI near the separatrix (see. fig \ref{sum}) and corresponding to these initial conditions: $\chi = 0$, $\phi_{c} = \pi$, $h= 0$ and $k=0$.

\begin{figure}[!h]
\centerline{\includegraphics[angle=0,width= 12cm,trim= 0 0 0 0]{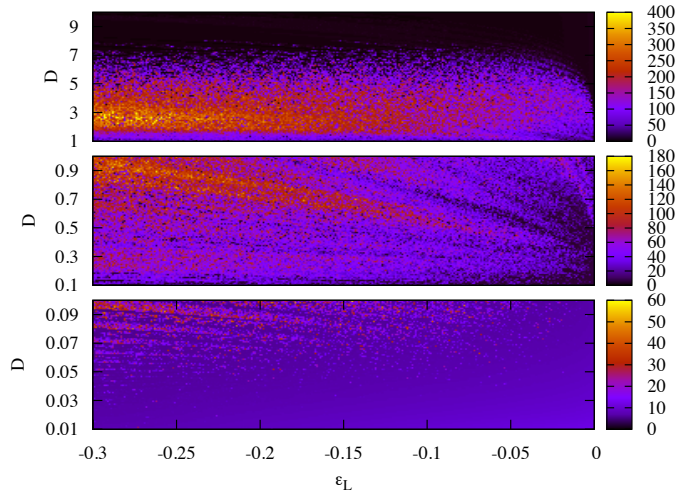}}    
\caption{%
Maps of FLI, for 90000 trajectories, the initial conditions used are : $\chi = 0$, $\phi_{c} = \pi$, $h= 0$ and $k=0$. 
We integrated numerically a several orbits ($90000$ orbits) during $2000$ years, with these initial conditions: $\chi = 0$, $\phi_{c} = \pi$, $h= 0$ and $k=0$, for 100 values of $\epsilon_{L}$ between $[-0.3,0]$ and 900 values of $D$ between $[0,7.5]$.
A FLI of 400 corresponds to 5 years in term of divergence of two close orbits, 
smaller values of the FLI correspond to longer divergence times that scale like 5 years/FLI.
}%
\label{fli1}
\end{figure}

The values of FLI represent the irregularity degree for each orbit. We show in figure (\ref{fli1}) a FLI map for several values of $D$ and $\epsilon_{L}$.
Here we are interested by the global dynamics of the test particle,
we note that for values of $D$ between $\sim 0$ and $2$, the chaoticity  area grows gradually (clear colors).
Beyond $D=6$, the orbits become regular (dark colors) what is in agreement with the previous discussion.
For $\epsilon_{L} =0$, the trajectories becomes regulars as expected, because the system is not perturbed as  the mass of the perturber satellite is equal to zero.


\subsection{Real applications}
\label{sat_saturn}

\begin{figure}[!h]
\centerline{\includegraphics[angle=0,totalheight=7cm,trim= 0 0 0 0]{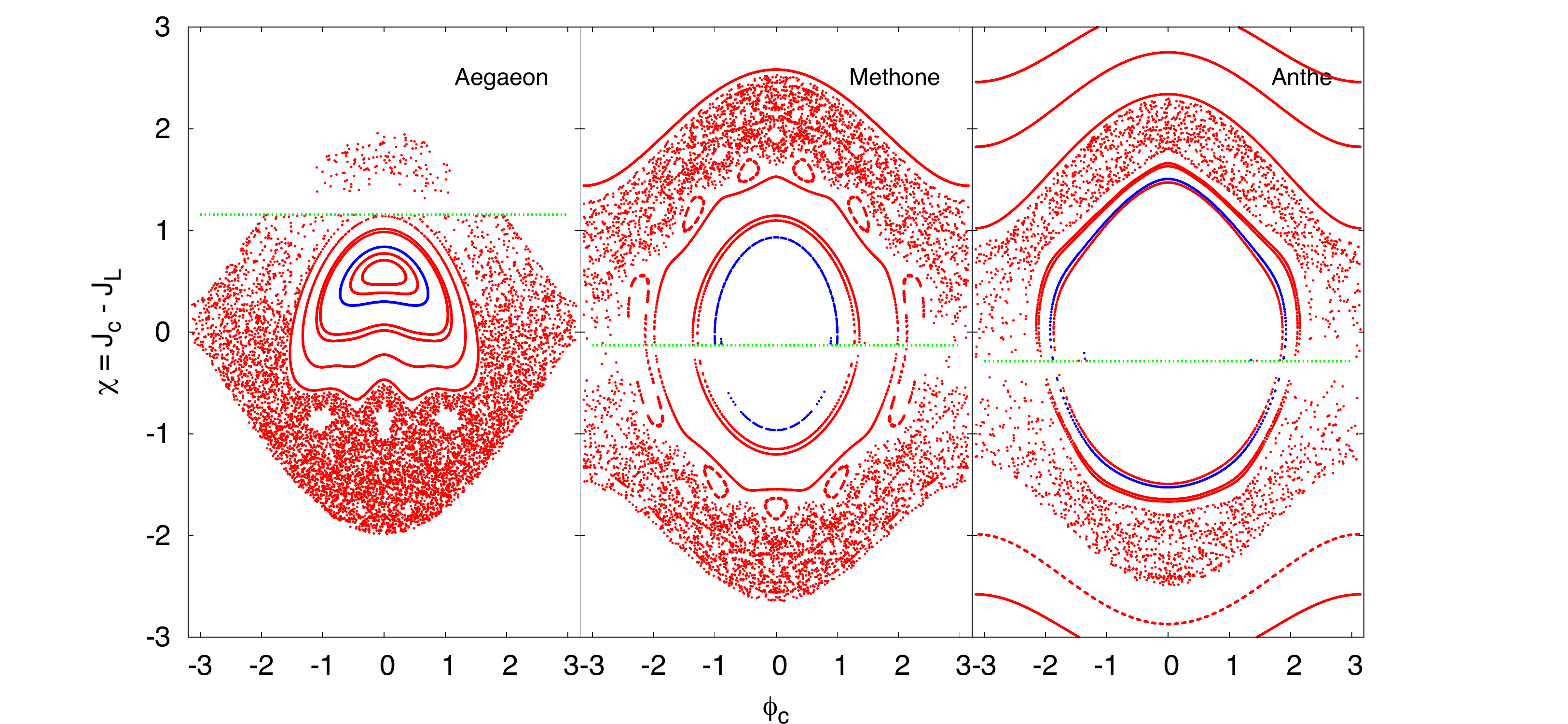}}    
\caption{%
Green curve : Position of Lindblad resonance at $J_c - J_L = -D$. Blue curve : Trajectory of the small satellite. Red curve : Surfaces of sections of a particle having the same parameters of each satellites with different initials conditions in term of $\chi = J_c - J_L$. Note that the corotation site is very affected by chaotocity due to the interaction with LER.
}%
\label{moons}
\end{figure}

In this section, we show that in some cases, a strong coupling between corotation and Lindbald resonances may lead to chaotic behavior. 
We apply the CoraLin model to several recently discovered small satellites dynamically linked to Mimas through first mean motion resonances : 
Anthe, Methone and Aegaeon (\textcolor{blue}{\cite{coo08,hed09,hed10}}), all associated with arc of material.
The presence of these structures are consistent with their confinement by CER with Mimas : 
Aegaeon is trapped in an inner 7:6 CER with Mimas, while Anthe and Methone are respectively near the outer 10:11 and a 14:15 CER resonances.
All satellites are trapped in CER with Mimas and perturbed enough ($D < 2$) by the associated LER.
Indeed, the topology of space phase depends of two parameters (table \ref{moons_param}) which are given by the configurations between the particle (one of these small moons) 
and Mimas (the disturbing satellite) orbiting a central planet (Saturn).
\begin{table}[h]
\centering
\begin{tabular}{|*{5}{c|}}
\hline
         & D  &  $\varepsilon_{L}$  & CER ($\phi_{c}$) & LER ($\phi_{L}$) \\
\hline
Aegaeon &     -1.155    &   0.132  &    $7\lambda' - 6\lambda - \varpi'$   & $7\lambda' - 6\lambda - \varpi$   \\
\hline
Methone &     0.129     & -0.115  &   $15\lambda - 14\lambda' - \varpi'$ & $15\lambda - 14\lambda' - \varpi$  \\
\hline
Anthe       &   0.286       &   -0.123 &  $11\lambda - 10\lambda' - \varpi'$  &  $11\lambda - 10\lambda' - \varpi$\\
\hline
\end{tabular}
\caption{
Critical angles and values of  $D$ and $\epsilon_L$ appropriates to Anthe, Methone and Aegaeon given by CoraLin Model, these parameters depend of the configurations of the small satellites in the Saturnian system.
We note that for the inner (outer) moons $D$ is  positive (negative) and $\epsilon_{L}$ is negative (positive). 
 }
\label{moons_param}
\end{table}
Poincar\'e surfaces of section reveal the dynamical structure of each orbit, and for some of them, their proximity to chaotic regions (fig \ref{moons}).
\\
Indeed the system~(\ref{eq_coralin0}) is numerically integrated for twenty separates trajectories and  ($J_{c} - J_{L}$) was plotted versus $\phi_{c}$.
From (fig \ref{moons}), we see that this dynamical system displays both regular and irregular orbits as most of the Hamiltonian systems.
A surface of section is obtained by getting back ($J_{c}, \phi_{c}$) with period $2\pi$ or only when $k=0$ and $\dot{k} > 0$.
This approach gives an idea on the structure of phase space for each of moons, we see also that the chaos results of a strong coupling between CER and LER, and that the chaotic sea is very large occupying totally the corotation site.
We notice that Anthe is close to a chaotic region. The amplitude of libration is $\pm 26$  km, i.e. about half of the corotation width ($50.7$ km). The orbit of Methone is farther away from the chaotic zone than Anthe, with an amplitude of libration of $\pm 20$ km and a corotation width of $55.36$ km. 
%
%
%
%
The case of Aegaeon (inside the orbit of Mimas) is slightly different, its orbit seems to be more stable than those of Anthe and Methone, 
as it is far enough from the chaotic region, the width of the corotation site is $31.62$ km, and the magnitude of libration in terms of semi-major axis is $\pm 4$ km  (\textcolor{blue}{\cite{hed09,hed10}}). 

We notice that the chaotic region is surrounding by many regular trajectories, this behavior means that the corotation stability is not affected, this result will be investigated in future works by analysis of capture probability of this structures inside CER site.\\
%
%
%


\section{Conclusion} 

We have studied the behavior of two satellites of masses $\mu_s$ and $\mu'_s$, 
revolving along coplanar, small-eccentricity orbits around a planet of mass $M_p \gg \mu_s, \mu'_s$. 
We have averaged the equations of motion close to a first order mean motion commensurability $m$+1:$m$,  
keeping only perturbing terms of first order in eccentricity 
and accounting for the secular apsidal precessions rates $\dot{\varpi}_s$ and $\dot{\varpi}'_s$ of the satellites.
This allows to derive the classical Hamiltonian of the system, with the two critical resonant arguments
$\phi =  (m+1)\lambda' - m\lambda - \varpi$ and
$\phi' =  (m+1)\lambda' - m\lambda - \varpi'$.
Using the constants of motion (total energy and angular momentum), 
the initial four-degree of freedom system reduces to 
a two-degree of freedom problem that is not in general integrable.

For $\dot{\varpi}_s- \dot{\varpi}'_s=0$ (as it is the case for the Keplerian problem), 
the Hamiltonian is integrable, a result initially derived by \textcolor{blue}{\cite{ses84}}.
The integrability of the system stems from the existence
of a second constant of motion, besides the Hamiltonian. 
Here we show that this constant is actually a generalized version of the Jacobi constant
(Eq.~(\ref{eq_jac_relat_gen})), 
where the orbital eccentricities of $\mu_s$ or $\mu'_s$ are replaced by a quantity that we call the
relative eccentricity of the two orbits (Eq.~(\ref{eq_e_relat})).
Secular terms forces a differential precession between the two orbits 
(i.e. $\dot{\varpi}_s- \dot{\varpi}'_s \neq 0$), 
and destroys the generalized Jacobi constant.

In the general case ($\mu_s \ne 0, \mu'_s \ne 0$), the two satellites play symmetrical roles.
This symmetry is broken, however, in the restricted problem (e.g. $\mu_s =0$).
The two critical resonant angles $\phi$ and $\phi'$,
or their counterparts  $\phi_L$ and $\phi_c$, 
are then associated with two resonances 
(Lindblad and corotation, or LER and CER, respectively) that plays very different roles. 
While the LER mainly excites the orbital eccentricity of the test particle,
leaving its semi-major axis relatively unaltered  (Eq.~(\ref{eq_eLin})),
the CER mainly changes its semi-major axis, leaving its eccentricity
almost constant (Eq.~(\ref{eq_ecor})).

The two resonances may be simultaneously described by a reduced Hamiltonian (Eq.~(\ref{eq_Hnorm})),
that depends upon only two dimensionless parameters that control the dynamics of the system: 
$(i)$ the distance $D \propto \dot{\varpi}_s - \dot{\varpi}'_s$ 
between the CER and LER, and 
$(ii)$ the forcing parameter $\epsilon_{L}$ that includes both the mass and the orbital eccentricity 
of the disturbing satellite, see Table~\ref{tab_eps_cn0}.

The resulting equations of motion are summarized by four simple differential equations that constitute
the ``CoraLin model" (see the system~(\ref{eq_coralin0})).
This system describes the coupling between the motion of a simple pendulum 
(Eq.~(\ref{eq_H_pend})) that has a separatrix of width $\pm2$,
and an Andoyer-type oscillator (Eq.~(\ref{eq_H_And})) centered at $\chi=-D$, see Fig.~\ref{fig_coralin}.
It has the advantage to permit a generic exploration of the dynamics of the system through simple numerical
integrations, Poincar\'e surface of section, etc...
Furthermore, it uses dimensionless parameters that encapsulate
all the parameters of the systems (mass and orbital eccentricity of the perturber, secular precessions of the orbits, etc...).

As an example, we have examined the phase portraits relevant to small Saturnian satellites trapped in CER's
with Mimas: Aegaeon, Methone and Anthe.
While the system is integrable for $D=0$, chaos is rapidly prevalent for values of $D$ as small as about 0.01, 
see Fig.~\ref{sum}. 
Only for large values of $D \gg 2$ is the system almost integrable again, 
with a behavior that can be qualitatively described using simple adiabatic invariant arguments.

More specific integrations (Fig.~\ref{moons}) show that  Aegaeon, Methone and especially Anthe are
close to prominent chaotic regions.
Future works are now in order to explain how those satellites may have been captured inside their
respective corotation sites. 
The numerical implementation of orbital migrations of Mimas and/or the small satellites in the CoraLin system
is in fact very simple, and we will use that model to explore various scenarii of resonant capture.
%
%
%
%
%
\begin{acknowledgements}
The authors thank Sylvio Ferraz-Mello for enlightening discussions concerning the integrability of the
3-body problem with two critical resonant arguments. 
We had several fruitful discussions with Philippe Robutel, Nicholas J. Cooper and Carl D. Murray,
and we thank Nicholas J. Cooper for providing orbital elements of Aegaeon, Methone and Anthe. 
We thank the Encelade working group for interesting discussions during regular meetings, in part funded by the EMERGENCE-UPMC project EME0911.
\end{acknowledgements}
%
%
%
%
%
%
%
%
%
%
%

\newcommand{\noopsort}[1]{}

%
%

%
%


\newpage

\nocite{*}

\bibliographystyle{icarus}      


\end{document}